\documentclass[twocolumn,showpacs,amsmath,amssymb,prl]{revtex4}

\usepackage{graphicx}
\usepackage{dcolumn}
\usepackage{bm}

\begin{document}
\title{Positron plasma diagnostics and temperature control
for antihydrogen production}

\author{
M.~Amoretti$^{1}$,
C.~Amsler$^2$, G.~Bonomi$^{3}$, A.~Bouchta$^3$, P. D.~Bowe$^4$,
C.~Carraro$^{1,5}$, C.~L. Cesar$^6$, M.~Charlton$^7$, M.~Doser$^3$,
V.~Filippini$^8$, A.~Fontana$^{8,9}$, M.~C. Fujiwara$^{10}$, 
R.~Funakoshi$^{10}$, P.~Genova$^{8,9}$, J.~S.~Hangst$^4$, 
R.~S.~Hayano$^{10}$, L.~V. J$\o$rgensen$^7$, V.~Lagomarsino$^{1,5}$,
R.~Landua$^3$, D.~Lindel\"{o}f$^2$, E.~Lodi~Rizzini$^{11}$, 
M.~Macr\'\i$^1$, N. Madsen$^2$, G. Manuzio$^{1,5}$, 
P. Montagna$^{8,9}$, H.~Pruys$^2$, C.~Regenfus$^2$, A.~Rotondi$^{8,9}$, 
G.~Testera$^{1,}$\footnote{Corresponding author\\Email address: gemma.testera@ge.infn.it}, A.~Variola$^1$, and D.~P.~van~der~Werf$^7$ \\
(ATHENA Collaboration) }

\affiliation{ 
$^1$Istituto Nazionale di Fisica Nucleare, Sezione di Genova, 16146 Genova, Italy\\
$^2$Physik-Institut, Z$\ddot{u}$rich University, CH-8057 Z\"{u}rich, Switzerland \\
$^3$EP Division, CERN, CH-1211 Geneva 23, Switzerland \\
$^4$Department of Physics and Astronomy, University of Aarhus, DK-8000 Aarhus C, Denmark \\
$^5$Dipartimento di Fisica, Universit\`{a} di Genova, 16146 Genova, Italy \\
$^6$Instituto de Fisica, Universidade Federal do Rio de Janeiro, Rio de Janeiro 21945-970, 
    and Centro Federal de Educa\c{c}\~{a}o Tecnologica do Ceara, Fortaleza 60040-531, Brazil \\
$^7$Department of Physics, University of Wales Swansea, Swansea SA2 8PP, UK \\
$^8$Istituto Nazionale di Fisica Nucleare, Sezione di Pavia, 27100 Pavia, Italy\\
$^9$Dipartimento di Fisica Nucleare e Teorica, Universit\`{a} di Pavia, 27100 Pavia, Italy \\
$^{10}$Department of Physics, University of Tokyo, Tokyo 113-0033, Japan \\
$^{11}$Dipartimento di Chimica e Fisica per l'Ingegneria e per i Materiali, Universit\`{a} di  
Brescia; Istituto Nazionale di Fisica Nucleare, Gruppo collegato di Brescia, 25123 Brescia, Italy 
}


\begin{abstract}
Production of antihydrogen atoms by mixing antiprotons with
a cold, confined, positron plasma depends critically on 
parameters such as the plasma density and temperature.  
We discuss non-destructive measurements, based on a novel, 
real-time analysis of excited, low-order plasma modes, that 
provide comprehensive characterization of the positron plasma 
in the ATHENA antihydrogen apparatus.  The plasma length, 
radius, density, and total particle number are obtained. 
Measurement and control of plasma temperature variations, 
and the application to antihydrogen production experiments 
are discussed.
\end{abstract}


\maketitle

The ATHENA collaboration recently produced and detected 
cold antihydrogen atoms~\cite{Amoretti2002a}
at the CERN Antiproton Decelerator (AD)~\cite{Maury1997}.
A similar result has been subsequently reported by the ATRAP 
collaboration ~\cite{Gabrielse2002}.
The antihydrogen was made by mixing low energy antiprotons with a cold, 
dense positron plasma in a nested Penning trap~\cite{Gabrielse1988}. 
A knowledge of the characteristics of the positron plasma is important
for several reasons. The most likely antihydrogen formation mechanisms,
spontaneous recombination and three body recombination,
have different dependences on both the density and the temperature
of the positron plasma~\cite{Holzscheiter1999}. Knowing these parameters 
is crucial in helping to elucidate the antihydrogen formation mechanism. 
Control and simultaneous monitoring of the positron plasma temperature allow 
the antihydrogen formation reaction to be effectively turned off 
while maintaining overlap between antiprotons and positrons.  
This provides a good measurement of the total background signal
for our unique antihydrogen detector~\cite{Amoretti2002a}.  
Furthermore, the space charge potential of a sufficiently dense
($10^{8}$ cm$^{-3}$) and extensive (length 3 cm) positron plasma  
considerably alters the effective electrostatic potential 
in the positron trap and thus the dynamics of the antiprotons in the nested trap.

Harmonically confined one component plasmas at temperatures close
to absolute zero are known to form spheroids of constant charge
density~\cite{Turner1987}.  In our case the shape is a prolate 
ellipsoid characterized by the aspect ratio $\alpha=z_p/r_p$ 
where $z_p$ and $r_p$ are the semi-major axis and 
semi-minor axis respectively [Fig.~1(a)].
A cold fluid theory~\cite{Dubin1991} relating the frequencies 
of the low-order plasma modes to the density and the aspect ratio 
of spheroidal plasmas was confirmed experimentally for laser cooled 
ion plasmas~\cite{Heinzen1991,Bollinger1993} and successfully applied
to cold electron plasmas~\cite{Weimer1994}. 
Work on finite temperature electron plasmas demonstrated that mode detection 
could be used as a diagnostic of density and aspect ratio and that for a  plasma
of known density and aspect ratio the frequency of the quadrupole mode 
is dependent on the plasma temperature~\cite{Tinkle,Higaki2002}. 

Here we describe an extension to the above work which provides a 
non-destructive diagnostic based on measurements  of the first 
two axial modes of a finite temperature positron plasma.
The diagnostic has no discernable effect on the normal evolution 
of the plasma, so it can be used during antihydrogen production.
We have developed a model in which the plasma length can be extracted 
from the shape of the resonance when the dipole mode is excited.  
Thus the aspect ratio, density, and length can be measured and the radius
 and positron number obtained.  We show that we can monitor induced changes 
in the temperature while ensuring that the normal evolution of other plasma 
parameters is not changed.  Using this monitor and suitable radio frequency 
excitation of the dipole mode we can set the plasma temperature during the 
interaction between  positrons and antiprotons.  

\begin{figure}
   \resizebox{0.45\textwidth}{!}{\includegraphics{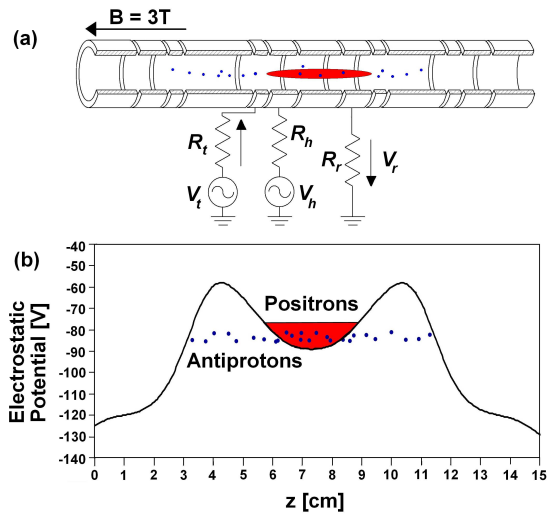}}
   \caption{ (a) Trap  electrodes with the heating and mode detection
    electronics. The shape of the positron plasma prolate ellipsoid is 
    shown schematically. (b) The axial potential of the ATHENA nested 
    trap is shown and the ranges of axial motion of the positrons and 
    the antiprotons indicated schematically.}
\end{figure}

In ATHENA, the positrons and antiprotons are confined 
by  cylindrical electrodes of radius $r_w$=1.25 cm. These are  inserted in 
a cryostat and immersed in a 3 T magnetic field. 
The positrons are held in the central part of the trap by an harmonic
axial potential [see Fig.~1(b)].
In the high magnetic field  the positrons  cool 
by synchrotron radiation~\cite{O'Neil1980}. 
We have not measured the temperature of the positron plasma but
note that the lower limit is set by the temperature of the trap electrodes 
(15 K). 

The first two  axial modes (dipole and quadrupole modes) are excited by
applying a sinusoidal perturbation to one electrode with an electromotive
force $V_t = v_t e^{j \omega t}$. The oscillation of the plasma induces 
a current in the pick-up electrode ~\cite{Kapetanakos1971,Wineland1975}
and a voltage $V_r = v_r(\omega)e^{j \omega t}$ is detected across the 
resistance $R_r$ [Fig.~1(b)]. Experimentally the ratio
$T_L(\omega)= v_r(\omega)/v_t$
is measured as a function of the drive frequency $\omega$ by
means of a network analyzer. A narrow step-wise frequency sweep is made 
of the voltage source across the resonance frequency of each mode.
The excitation amplitude $v_t$ is of the order of 100 $\mu$V and the dwell
time for each 5~kHz step is 3.96~ms. The number of scan steps is 
usually 50.  The choice of scan width is a balance between following 
scan-to-scan changes in mode frequency and avoiding perturbations to the
normal evolution of the plasma.                                     
To follow changes in the plasma over a longer time scale an automatic 
frequency tracking code has been implemented to allow the excitation to be
locked to the mode frequencies. For each frequency step, the amplitude and
phase  (relative to that of the driving signal) of the voltage induced by
the plasma motion are acquired. The cross talk signal between the 
transmitting and receiving electrodes is acquired without positrons and 
subtracted from the signal measured with the plasma present. 

The plasma number density $n$ and aspect ratio $\alpha$ can be extracted 
from the zero-temperature analytical model~\cite{Dubin1991} 
 using the measured 
frequencies ($\omega_1$, $\omega_2$) of the dipole and quadrupole modes.  
We have developed an equivalent circuit model which explicitely 
includes the plasma dimensions.
This method yields directly the plasma length when $\alpha$ and $n$ are known.
In contrast other equivalent circuit approaches utilizing tuned circuits
and assuming small cloud dimensions measured the total number of
particles in Penning traps ~\cite{Wineland1975}. Our model
describes the signal induced on an electrode by the coherent oscillations
of the dipole mode when an external driving force is applied.
In particular we 
write~\cite{AmorettiSubmitted}
\begin{equation}\label{curve}
T_{L} = \frac{g_t(\alpha,z_p) g_r(\alpha,z_p) R_r }
{R_s + j\omega L \left(1-\omega_{1}^2/\omega^2\right)}.
\end{equation}
The dimensionless functions $g_r$ and $g_t$ 
depend on the shape of the plasma
and on the geometry of the trap.
They describe the effects of the finite
plasma extension both on the mode excitation and detection.
$g_t(\alpha,z_p) V_t/2 r_w$ is the  electric field
acting on the center of mass of the plasma when the dipole 
mode is excited by applying a voltage $V_t$ to the transmitting
electrode with the other electrodes  grounded.
 The function $g_r(\alpha,z_p)$ is related to the
current $I_r$ induced  by the dipole mode oscillation on the receiving 
electrode, $I_r=N e g_r(\alpha,z_p)v_{cm}/2 r_w$ where $v_{cm}$
is the velocity of the particles due to the dipole mode and $e$ is the
positron (electron) charge.
$T_L$ is obtained by measuring $T_L^{\prime}=A T_L$ where 
$A$ is the net gain of the electronics chain and it is
 independent of the plasma properties.
We used electrons to fine tune this parameter
by comparing the number obtained by this diagnostic with the number measured
on a Faraday cup.
Electrons were chosen as they can be loaded faster, with a wider range 
of total number N. The inductance $L$ of the equivalent circuit is 
related to the plasma length,
\begin{equation}
L = \frac {3 \alpha^2 r_w^2 m}{\pi n e^2 z_p^3},
\end{equation}
where  $m$ is the positron (or electron) mass ~\cite{AmorettiSubmitted}.
The resistance $R_s$ characterizes the damping rate of the mode. 
$\alpha$ and $n$ are determinated independently by the frequency analysis. 
The power transmitted trough the
plasma $|T_L|^2$ is related to $R_s$ and $z_p$ by Eq.~(2) and Eq.~(1).
Thus a fit to the measured transmitted power 
yields $z_p$. The radius $r_p$ and the total number $N$ can now be found. 
\begin{figure}
    \resizebox{0.40\textwidth}{!}{\includegraphics{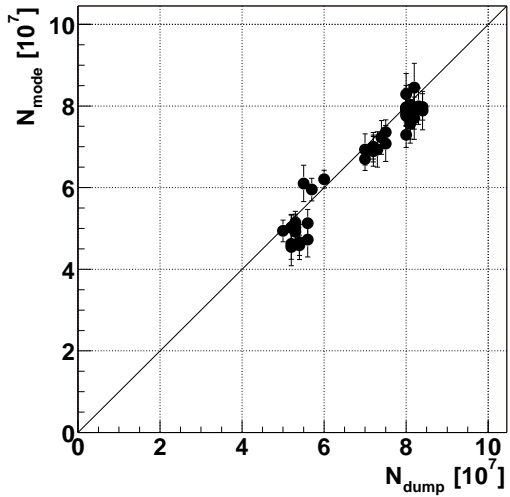}}
    \caption{The total number of positrons obtained using the modes 
    diagnostic is plotted against the number obtained by extracting
    the positrons to a Faraday cup.}
\end{figure}
An example of an application of this diagnostic to positrons 
is shown in Fig.~2. Here the total number obtained with the model 
is plotted against the number found by extraction to a Faraday cup. 
In this regime the linearity and good correspondence in the absolute
number show that both the model and its implementation constitute 
a complete, real-time, non-destructive plasma diagnostic. 
Typical properties of the ATHENA positron plasma for antihydrogen production
were $7 \times 10^{7}$ positrons at a density of about 
$1.7 \times 10^{8}$ cm$^{-3}$ in a plasma approx. 3.2 cm long ($2z_p$)
with a radius of about 0.25 cm and a storage time of several hundreds
of seconds. The maximum change in the plasma parameters during the 
mixing cycle of 190 seconds was less than 10\%. 

\begin{figure}[b]
    \resizebox{0.40\textwidth}{!}{\includegraphics{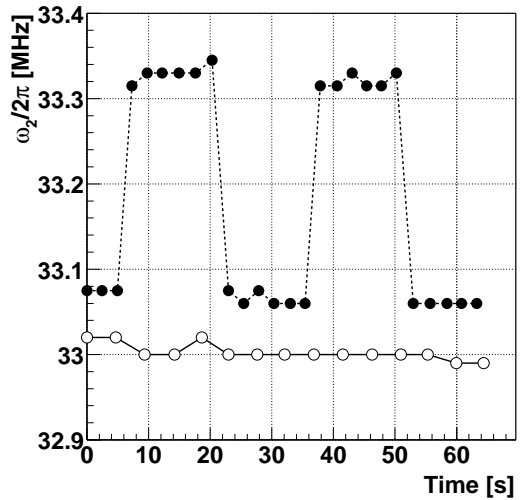}}
    \caption{The  quadrupole mode frequency versus time for normal 
    evolution ($\circ$) and for two heat off-on cycles ($\bullet$). 
    The frequency shift corresponds to an increase of the plasma 
    temperature of about 150 meV.}
\end{figure}

Because of the importance of temperature in antihydrogen production,
we have also investigated whether this diagnostic system can be used
to detect changes in the plasma temperature. 
Previous authors have demonstrated, with work on similar electron
plasmas~\cite{Tinkle,Higaki2002}, that temperature changes
manifest themselves in frequency shifts of the quadrupole mode. 
In particular they find
\begin{equation}\label{tempmodel}
k\Delta T = \frac{m z_p^2}{5}  [(\omega_{2}^h)^2 - (\omega_2)^2] 
\left[3 - \frac{\omega_p^2 \alpha^2}{2\omega_2^2} 
\frac{d^2f(\alpha)}{d\alpha^2} \right]^{-1}, 
\end{equation}
where $f(\alpha) = 2 Q_1(\alpha/\sqrt{\alpha^2-1})/(\alpha^2-1)$ and $Q_1$ 
is a Legendre function of the second kind, $\omega_{2}^h$ is 
the quadrupole frequency with heating applied and $\omega_p$
is the plasma frequency. The values of $\alpha$, $\omega_p$ and $z_p$ 
are evaluated in the cold fluid limit. Provided that these parameters 
do not vary significantly when the plasma temperature changes, 
a measured shift in the quadrupole frequency can be used 
to calculate the magnitude of the temperature change.

\begin{figure*}[t]
    \resizebox{0.8\textwidth}{!}{\includegraphics{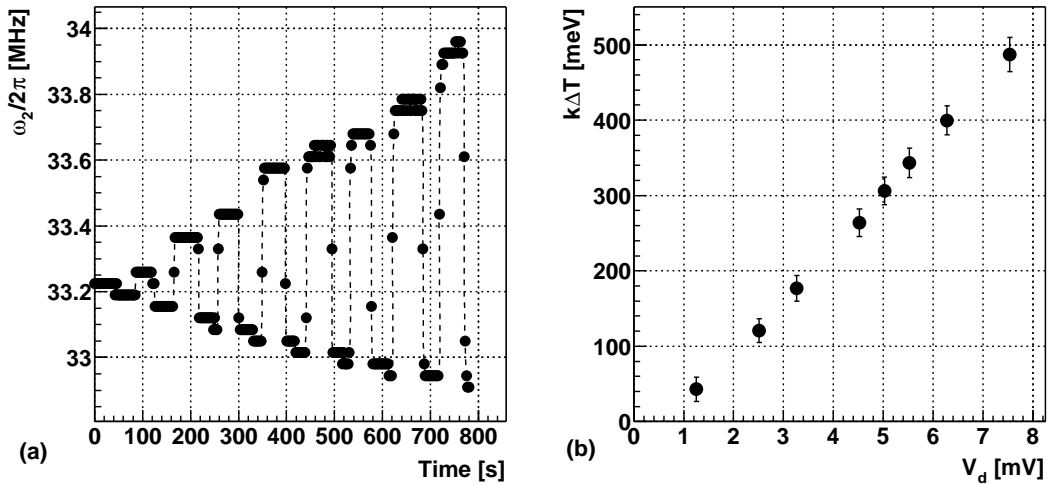}}
    \caption{(a) Time evolution of the quadrupole mode frequency during a 
heating off-on cycle. The different frequency shifts correspond to 
different heating amplitudes $V_d$.
The drift of the frequency in the unperturbed intervals is due to
the plasma expansion and it is consistent with the normal evolution
of the unperturbed plasma (see also Fig.~3).
(b) Dependence of the temperature variation on radio-frequency 
signal amplitude for the same cycle. }
\end{figure*}

We have used the mode diagnostic system to investigate the response
of a cold and dense positron plasma without antiprotons to heating, 
implemented by applying an excitation near the dipole frequency 
(21 MHz) to one of the trap electrodes [Fig.~1(a)].  Off-resonance
heating pulses were not effective. 
The excitation is a variable amplitude signal that is swept from 
20 MHz to 22 MHz at a repetition rate of about 1~kHz.
This is done to ensure that the dipole mode frequency 
is covered and that the plasma 
reaches thermal equilibrium. 
Previous authors report an equilibration rate of some tens of kHz 
in similar conditions \cite{Beck}.
Figure 3 shows the behavior of the quadrupole frequency when the
initially cold plasma is subjected to heating off/heating on cycles. 
Application of the excitation results in a rapid, voltage-dependent 
rise in the quadrupole frequency.  When the excitation is removed, the
quadrupole frequency returns to a value in step with the evolution of
the unperturbed plasma which is also shown in Fig.~3.
The unperturbed plasma evolution is characterized by a slow decrease
in the frequency of the quadrupole mode and corresponding decrease in 
aspect ratio and density. This is consistent with a slow expansion of the
plasma. 
 The heating
and the cooling times are faster than our sampling interval resulting in the
observed 
discontinuities of the frequency
shifts on the time scale of Fig.~3. 

Furthermore, for these heating amplitudes, without antiprotons we 
observe no discernable positron loss with our positron annihilation
detector which covers a solid angle of ~80$\%$ of 4$\pi$ \cite{Rainfoot}.
Figure~4(a) shows the response of the plasma quadrupole frequency 
during heating off/heating on cycles with different amplitudes $V_d$ 
of generated heating voltage [directly proportional to $V_h$ of Fig.~1(a)].
Thus the dependence of temperature increase $\Delta T$ on $V_d$ was 
found and the  result is shown in Fig.~4(b).
For the data reported here the minimum measureable temperature was about 
15 meV due to the frequency step size used (20 kHz). 
The minimum step size of 5 kHz would give a sensitivity of a few meV. 
The observed linear dependence of the temperature rise on $V_d$,
is in contrast with the linear dependance on the the power (and thus on
$V_d^2$) that one would expect. This could be  most likely due to
non linear effects of the on-resonance heating. 
 A nonlinear regime could also explain the fact
that the temperature rise in Fig.~4(b) appears to extrapolate to zero
at finite  $V_d$.

The utility of the mode diagnostic system as applied to antihydrogen
production is immediately apparent if we consider that the radiative
reaction rate is proportional to $n$ and to $T^{-1/2}$, while the 
three-body reaction rate varies as $n^2$ 
and $T^{-9/2}$~\cite{Holzscheiter1999}.
In order to gain insight into the experimental production mechanism
and to ensure reproducible results, careful monitoring of any changes
in these parameters is essential. 
The ability to reproducibly add heat to the positron plasma, and to
thereby control the reaction rate, is also very desirable.  We have 
already utilized the heating and temperature diagnostic to \lq\lq turn
off\rq\rq  antihydrogen production in ATHENA~\cite{Amoretti2002a}, 
providing a useful null measurement. 

The implementation of automated mode analysis measurement has allowed
us to log the plasma parameters at all times when positrons are in
the trap. The system is wide band and does not require that any 
resonant circuitry be mounted internally on trap electrodes. We are 
thus free to choose a wide range of depths for our harmonic positron
trap and are therefore able to tune the shape of our positron plasma
at will.  Knowing the size of the positron cloud is crucial to assuring
a good spatial overlap between antiprotons and positrons in the 
reaction region. 
The  space charge potential of our positron plasma flattens the potential
inside its volume along the z direction reducing the potential barrier 
seen by the antiprotons  in the mixing trap by several volts. 
Information about the effective potential is important in helping to 
understand the antiproton cooling and interaction dynamics.  

We have extended the plasma mode diagnostic method to provide comprehensive 
characterization of a cold, dense positron plasma employed in the ATHENA 
antihydrogen experiment. The method has already been utilized to great 
advantage in ATHENA, and promises to be an essential element of future 
experiments.  
The technique, while particularly useful for non-destructive measurements
on difficult-to-produce species such as positrons, has immediate 
applicability to other Penning trap plasmas.

This work was supported by INFN (Italy),CNPq and FAPERJ
(Brazil), MEXT 
(Japan), SNF (Switzerland), SNF (Denmark) and EPSRC (UK). 


\end{document}